%% file: main.tex
\documentclass[conference,letterpaper]{IEEEtran}

\usepackage{graphicx}
\usepackage{epsfig}
\usepackage{epstopdf}
\usepackage{amsmath,amsthm,amsfonts,amssymb}
\usepackage{bbm}
\usepackage{multirow}
\usepackage{cite}
\usepackage{tikz,pgfplots}
\usepackage{tikz-cd}
\usepackage{ellipsis}
\usepackage{xcolor}
\usepackage{colortbl} 
\usepackage{xifthen}
\usepackage{balance}
\usepackage{subcaption}
\usepackage{cancel}
\usepackage{xfrac}
\usepackage[commandnameprefix=ifneeded,commentmarkup=uwave]{changes}
\definechangesauthor[name=Pascal, color=blue]{P}
\definechangesauthor[name=Valerio, color=purple]{V}
\definechangesauthor[name=Charles, color=red]{C}
\usetikzlibrary{patterns,arrows.meta}
\usepackage{glossaries} 
\usepackage[draft,pdfborder={0 0 0}]{hyperref}

\newtheorem{proposition}{Proposition}
\theoremstyle{definition}
\newtheorem{theorem}{Theorem}
\theoremstyle{definition}

\theoremstyle{definition}

\theoremstyle{definition}
\newtheorem{definition}{Definition}
\theoremstyle{definition}

\theoremstyle{remark}

\DeclareMathOperator{\dec}{dec}
\DeclareMathOperator{\adec}{adec}
\DeclareMathOperator{\eval}{eval}

\newcommand{\degtxt}{\text{deg}}

\newcommand{\pid}{\mathbbm{1}}
\newcommand{\Ical}{\mathcal{I}}
\newcommand{\Fcal}{\mathcal{F}}
\newcommand{\Ccal}{\mathcal{C}}
\newcommand{\Acal}{\mathcal{A}}
\newcommand{\Gcal}{\mathcal{G}}

\newcommand{\GA}{\mathsf{GA}}

\newcommand{\BLTA}{\mathsf{BLTA}}

\newcommand{\Aut}{\mathsf{Aut}}

\newcommand{\Sbf}{\mathbf{S}}
\newcommand{\ybf}{\mathbf{y}}
\newcommand{\xbf}{\mathbf{x}}
\newcommand{\ubf}{\mathbf{u}}
\newcommand{\Vbar}{\overline{V}}
\newcommand{\mbar}{\overline{m}}

\definecolor{matlab1}{rgb}{0.000, 0.447, 0.741} 
\definecolor{matlab2}{rgb}{0.850, 0.325, 0.098} 
\definecolor{matlab3}{rgb}{0.929, 0.694, 0.125} 
\definecolor{matlab4}{rgb}{0.494, 0.184, 0.556} 
\definecolor{matlab5}{rgb}{0.466, 0.674, 0.188} 
\definecolor{matlab6}{rgb}{0.301, 0.745, 0.933} 
\definecolor{matlab7}{rgb}{0.635, 0.078, 0.184} 

\begin{document}
\newacronym{rmpsc}{PS-RM}{Partially-Symmetric Reed-Muller}
\newacronym{upo}{UPO}{universal partial order}
\newacronym{ae}{AE}{Automorphism Ensemble}
\newacronym{rm}{RM}{Reed-Muller}
\newacronym{psc}{PS}{Partially-Symmetric}
\newacronym{scl}{SCL}{Successive-Cancellation List}
\newacronym{ml}{ML}{Maximum-Likelihood}
\newacronym{sc}{SC}{Successive-Cancellation}
\newacronym{bp}{BP}{Belief Propagation}
\newacronym{scan}{SCAN}{Soft Cancellation}

\newacronym{cascl}{CA-SCL}{CRC-aided Successive Cancellation List}
\title{On the Distribution of Partially-Symmetric Codes for Automorphism Ensemble Decoding}

\author{\IEEEauthorblockN{Charles Pillet\IEEEauthorrefmark{1}, Valerio Bioglio\IEEEauthorrefmark{2}, and Pascal Giard\IEEEauthorrefmark{1}}
  \IEEEauthorblockA{\IEEEauthorrefmark{1}Department of Electrical Engineering, \'Ecole de technologie sup\'erieure, Montr\'eal, Qu\'ebec, Canada.}
  \IEEEauthorblockA{\IEEEauthorrefmark{2}Department of Computer Science, Università degli Studi di Torino.}
  Email: charles.pillet.1@ens.etsmtl.ca, valerio.bioglio@unito.it, pascal.giard@etsmtl.ca}%

\maketitle

\begin{abstract}
\Gls{ae} decoding has recently drawn attention as a possible alternative to list decoding of polar codes.
In this letter, we investigate the distribution of \gls{rmpsc} codes, a family of polar codes yielding good performances under \gls{ae} decoding.
We prove the existence of these codes for almost all code dimensions for code lengths $N\leq 256$. 
Moreover, we analyze the absorption group of this family of codes under SC decoding, proving that valuable permutations in \gls{ae} decoding always exist. 
Finally, we experimentally show that \gls{rmpsc} codes can outperform state-of-the-art polar-code-construction algorithms in terms of error-correction performance for short code lengths, while reducing decoding latency. 
\end{abstract}
\glsresetall
\section{Introduction}
\label{sec:intro}
Polar codes, introduced in 2009 \cite{ArikanFirst}, have been adopted in the $5^{th}$ generation of mobile networks (5G) \cite{polar_5G} in late 2016. 
During the standardization process, it was agreed that \gls{cascl} \cite{TalSCL} would be adopted as the standard decoding algorithm \cite{3GPP_TS}. 
\gls{cascl} is hard output, limiting its use in iterative receivers, moreover its sequential schedule coupled with comparisons of up to $2L$ candidate codewords at each estimated information bit increase its decoding latency.

Recently, \emph{ensemble decoding} was proposed for polar codes \cite{Perm_auto}.
Ensemble decoding offers various decoding possibilities such as both soft or hard output, and parallel or serial implementation.
In ensemble decoding, $M$ instances of the same polar decoder are run independently for different permuted versions of the received signal. That decoder could be, for example, the hard-output \gls{sc} decoder \cite{Perm_auto} or the soft-output decoders \gls{bp} \cite{BPL} or \gls{scan} \cite{SCANL}.
Preliminary hardware-implementation results showed that ensemble decoding with \gls{bp} can outperform other variants of \gls{bp} decoding both in terms of throughput and area efficiency \cite{BPL_HW}.

Very recently, polar code design for ensemble decoding has been gaining momentum, e.g., \cite{Perm_auto,Stuttgart_AE_RM,Stuttgart_AE_PC,PSMC,Paris_AE_v1,Perm_auto_subcode,Paris_AE_v2}.
The use of affine automorphisms effectively improves decoding performance of ensemble decoding. The resulting decoder is labeled as an \gls{ae} decoder in \cite{Stuttgart_AE_RM}. 
However, \gls{ae} decoding does not seem to be as effective for long codes.
Authors in \cite{PSMC} introduce \gls{psc} codes to prove that $M$ should grow exponentially with the code length to get good error-correction performance, while reliability-based design only provides a limited affine automorphism group when the code length grows \cite{few_auto}. 
On the other hand, \gls{psc} codes, known to have an advantageous affine automorphism group, exhibit good performance under \gls{ae} decoding for practical decoding parameters \cite{Stuttgart_AE_PC,Paris_AE_v1}.
The output of \gls{ae}-\gls{sc} decoding permits to cluster the automorphisms into groups providing the same codeword candidate \cite{Paris_AE_v2}.
The number of these groups, also called equivalence classes, gives an upper bound on $M$. 

To yield good performance under \gls{ae} decoding, two criteria are crucial in the design of short polar codes.
First, the minimum distance of the code should be maximized to avoid poor performance under \gls{ml} decoding \cite{BPL}.
Second, the code should be partially or fully symmetric, exhibiting a meaningful affine automorphism group for \gls{ae} decoding \cite{Stuttgart_AE_PC,PSMC,Paris_AE_v1}.
In this paper, the feasibility of these high-performance codes, termed \gls{rmpsc} codes in the remainder, is investigated for code lengths $N\leq 256$.
We prove that the absorption group of a \gls{rmpsc} code has a promising structure for \gls{ae} decoding, allowing better error-correction performance with reduced decoding latency with respect to short 5G polar codes under \gls{cascl} decoding. 

\section{Preliminaries}\label{sec:preliminaries}
\subsection{Polar Codes}
A $(N=2^n,K)$ polar code of length $N$ and dimension $K$ is a binary block code defined by a kernel $\mathbf{T}_2\triangleq \left[ \begin{smallmatrix} 1 & 0\\ 1 & 1 \end{smallmatrix} \right]$, the transformation matrix $\mathbf{T}_N=\mathbf{T}_2^{\otimes n}$, the information set $\Ical\subseteq[N]$ with $[N]=\{0,1,\dots,N-1\}$ and the frozen set $\Fcal=[N]\setminus\Ical$.
The encoding is performed as $\xbf=\ubf\cdot \mathbf{T}_N$ with the input vector $\ubf=\left(u_0,\dots,u_{N-1}\right)$ generated by assigning $u_i=0$ if $i\in\Fcal$ and storing information in the $K$ bit-channels stated in $\Ical$.
In polar coding, $\Ical$  includes the $K$ most reliable bit channels resulting from channel polarization \cite{ArikanFirst}.
If the $K$ most reliable channels are chosen among the ones composing a larger \gls{rm} code, the resulting code distance is maximized and the code is referred to as a \emph{RM-polar} \cite{RM_polar}.
RM-polar codes have a lower \gls{ml} bound and show enhanced performance under ensemble decoding \cite{BPL}.

The structure of $\mathbf{T}_N$ permits to define a \gls{upo} $j\preceq i$ between two virtual channels $i,j\in[N]$ that is valid for every communication channel \cite{PartialOrder}.
A polar code defined by $\Ical$ is compliant with the \gls{upo} if $\forall i,j\in[N]$ with $i\succeq j$, $j\in\Ical$ implies that $i\in\Ical$.
A polar code fulfilling the \gls{upo} can be described through few generators composing the minimum information set $\Ical_{min}^N\subseteq[N]$ \cite{Stuttgart_AE_PC} as
\begin{align}\label{eq:Imin}
\Ical=\bigcup_{j\in\Ical_{min}^N} \{i\in[N],j\preceq i\}.
\end{align}
In the following, a code that maximizes the code distance and that is compliant with the \gls{upo} will also be referred to as a RM-polar by a slight abuse of notation.
\subsection{Monomial Codes}
Let us define $\mathcal{M}^{[n]}$ as the set of monomials in $n$ variables over $\mathbb{F}_2$. 
A monomial $m\in\mathcal{M}^{[n]}$ is defined as a product of the variables $\{V_0,\dots,V_{n-1}\}$ and its degree $\degtxt(m)$ corresponds to the number of variables in its product.
Similarly, a negative monomial $\mbar$ is defined as a product of the variables $\{\Vbar_0,\dots,\Vbar_{n-1}\}$ with $\Vbar_i=1\oplus V_i$.
A monomial $\mbar_l\in\mathcal{M}^{[n]}$ is connected to the integer $l\in [N]$  as $\mbar_l=\prod_{i\notin S}\Vbar_i$, with $l=\sum_{i\in S}2^i$, $S\subseteq[n]$.
We have $\degtxt(\mbar_l)=n-|S|$ and there are $\binom{n}{k}$ degree-$k$ monomials.
A boolean function $f\,:\,\mathbb{F}_2^n\rightarrow\mathbb{F}_2$ can be seen as a map associating a bit to each integer in $[N]$.
The evaluation function $\eval\,:\,\mathbb{F}_2^n\rightarrow\mathbb{F}_2^N$ checks the output of $f$ for all elements of $\mathbb{F}_2^n$ in increasing order, namely:
\begin{align}
\eval(f) = (f(\mathbf{b}_0), f(\mathbf{b}_1), . . . , f(\mathbf{b}_{N-1}),
\end{align}
with $\mathbf{b}_k\in\mathbb{F}_2^n$ being the binary representation of  $k \in[N]$.
Monomial codes of length $N=2^n$ are a family of codes that can be obtained as evaluations of monomials in $n$ binary variables. 
A monomial code $\Ccal(N,K,\Gcal)$ is generated by $K$ monomials forming the \emph{generating monomial set} $\Gcal$ of the code.
Polar codes can be described through this formalism \cite{BardetPolyPC} since $\forall l\in[N],\, \eval(\mbar_l)$ correspond to the $l^{th}$ row of $\mathbf{T}_N$ \cite{BardetPolyPC}.
The notion of partial order also exists among the monomials, namely:
\begin{definition}[Partial ordering]\label{def:PO}
Given two monomials $\mbar_{t_1},\mbar_{t_2}$ of degrees $s_1$ and $s_2$, where $\mbar_{t_1}=\Vbar_{i_0} \cdot \ldots \cdot \Vbar_{i_{s_1-1}}$, with $i_0<\dots<i_{s_1-1}$ and $\mbar_{t_2}=\Vbar_{j_0} \cdot \ldots \cdot \Vbar_{j_{s_2-1}}$, with $j_0<\dots<j_{s_2-1}$, we say that $\mbar_{t_1} \preceq \mbar_{t_2}$ in two cases: 
if $s_1=s_2=s$, when $i_l \leq j_l$ for all $l=0,\dots,s-1$, while if $s_1<s_2$, when there exists a divisor $\mbar_{t'}$ of $\mbar_{t_2}$ with $\degtxt(\mbar_{t_1}) = \degtxt(\mbar_{t'})$ and $\mbar_{t_1} \preceq \mbar_{t'}$. 
\end{definition}
A monomial code is called \emph{decreasing} if $\forall\, \mbar_{t_1},\mbar_{t_2}\in\mathcal{M}^{[n]}$ such that $\mbar_{t_1}\preceq \mbar_{t_2}$, if $\mbar_{t_2}\in\Gcal$, then $\mbar_{t_1}\in\Gcal$. 
In this case, $\Gcal$ can be also described with the minimal monomial set $\Gcal_{min}^N$ containing the few monomials necessary to generate  $\Gcal$ 
\begin{align}
\Gcal=\bigcup_{\mbar_j\in\Gcal_{min}^N} \{\mbar_i\in\mathcal{M}^{[N]},\mbar_i\preceq \mbar_j\}.
\end{align}
A polar code compliant with the \gls{upo} framework is a provably decreasing monomial \cite{BardetPolyPC}. 
Thus, $\Ical_{min}^N$ and $\Gcal_{min}^N$ generate the same code. 
Next, we use the notation $\Ical_{min}^N$ for both sets. 

\subsection{Automorphism Group of Polar Codes}\label{subsec:auto_group_pc}
Let $\Pi(N)$ be the set of all permutations on $\{0,\dots,N-1\}$.
An automorphism $\pi\in\Pi(N)$ of a code $\Ccal$ is a permutation of $N$ elements mapping every codeword $\xbf \in \Ccal$ into another codeword $\pi(\xbf) \in \Ccal$. 
The automorphism group $\Aut(\Ccal)$ of a code $\Ccal$ is the set containing all automorphisms of the code. 
For monomial codes, the \emph{affine automorphism group} $\Acal \subseteq \Aut(\Ccal)$, formed by the automorphism that can be written as affine transformations of $n$ variables, is of particular interest. 
An affine transformation of $n$ variables is described by 
\begin{align}
	\mathbf{z} & \mapsto \mathbf{z'} = \mathbf{A} \mathbf{z} + \mathbf{b} ,
	\label{eq:GA}
\end{align}
where $\mathbf{z} , \mathbf{z'} \in \mathbb{F}_2^n$, the matrix $\mathbf{A}\in\mathbb{F}_2^{n\times n}$ is invertible and $\mathbf{b}\in\mathbb{F}_2^{n}$. 
The variables in \eqref{eq:GA} are the binary representations of code bit indices, and thus affine transformations represent code bit permutations.
The automorphism group of any RM code with order $r$ and variable $n$, denoted $\mathcal{R}(r,n)$, is known to be the complete affine group $\GA(n)$ \cite{WilliamsSloane}.
The affine automorphism group of codes compliant with the \gls{upo} is the block-lower-triangular affine (BLTA) group \cite{Stuttgart_AE_PC,li2021complete}, namely the group of affine transformations having a BLT transformation matrix. 
The $\BLTA$ group is recovered by the sizes of the blocks alongside the diagonal, defining the profile $\Sbf=(s_1,\ldots,s_l)$, with $s_1+\ldots+s_l=n$. 

\subsection{AE Decoder, Partial Derivatives and Code Symmetry}
Given a decoder $\dec$ for a code $\Ccal$, the corresponding automorphism decoder $\adec$ is given by
\begin{equation}
\label{eq:adec}
\adec(\ybf,\pi) = \pi^{-1}\left( \dec(\pi(\ybf)) \right),
\end{equation}
where $\ybf$ is the received signal and $\pi,\pi^{-1} \in \Aut(\Ccal)$. 
The \gls{ae} decoder \cite{Stuttgart_AE_RM} consists of $M$ $\adec$ instances running in parallel with $\pi\in\Acal$.
The codeword candidate of \gls{ae} is selected using a least-squares metric. 
The absorption group of $\dec$, denoted $[\pid]$, gathers all permutations such that $\forall \ybf\in\mathbb{R}^N,\,\adec(\ybf,\pi) = \dec(\ybf)$ \cite{Paris_AE_v2}.
The absorption group of SC $[\pid]\subseteq \Acal$ is also a $\BLTA$ group \cite{SC_invariant}. 
It is possible to use $[\pid]$ to cluster $\Acal$ into $\sfrac{|\Acal|}{[\pid]}$ equivalence classes, namely subsets of automorphisms always providing same results under \gls{sc} \cite{Paris_AE_v2}.

Given a  monomial code $\Ccal(N,K,\Gcal)$, its partial derivative $\frac{\partial}{\partial \Vbar_i}$ is the monomial code $\Ccal_i\left(\frac{N}{2},K_i,\Gcal_i\right)$.
$\Gcal_i$ is composed of all monomials in $\Gcal$ including the variable $\Vbar_i$. 
By definition, there exist $n$ partial derivatives $\Ccal_0,\dots,\Ccal_{n-1}$, of dimensions $K_0,\dots,K_{n-1}$.
As an example, the partial derivatives 
of the code $\Ccal\left(8,4,\Gcal=\left\{1,\Vbar_0,\Vbar_1,\Vbar_0\Vbar_1\right\}\right)$ are calculated as: 

{\footnotesize
\begin{align*}
    \dfrac{\partial}{\partial \Vbar_0}&=\left\{\dfrac{\partial 1}{\partial \Vbar_0}=0,\dfrac{\partial \Vbar_0}{\partial \Vbar_0}=1,\dfrac{\partial \Vbar_1}{\partial \Vbar_0}=0,\dfrac{\partial \Vbar_0\Vbar_1}{\partial \Vbar_0}=\Vbar_1\right\}=\left\{1,\Vbar_1\right\},\\
    \dfrac{\partial}{\partial \Vbar_1}&=\left\{\dfrac{\partial 1}{\partial \Vbar_1}=0,\dfrac{\partial \Vbar_0}{\partial \Vbar_1}=0,\dfrac{\partial \Vbar_1}{\partial \Vbar_1}=1 ,\dfrac{\partial \Vbar_0\Vbar_1}{\partial \Vbar_1}=\Vbar_0\right\}=\left\{1,\Vbar_0\right\},\\
    \dfrac{\partial}{\partial \Vbar_2}&=\left\{\dfrac{\partial 1}{\partial \Vbar_2}=0,\dfrac{\partial \Vbar_0}{\partial \Vbar_2}=0,\dfrac{\partial \Vbar_1}{\partial \Vbar_2}=0,\dfrac{\partial \Vbar_0\Vbar_1}{\partial \Vbar_2}=0\right\}=\emptyset.
\end{align*}
}%
The code symmetry $t$ is defined as the number of partial derivatives having the lowest dimension \cite{PSMC} and coincides with the last block of size $s_l\times s_l$ of $\Acal=\BLTA(s_1,\dots,s_l=t)$ \cite{Stuttgart_AE_PC,Paris_AE_v2}.
Having $t\neq 1$ for this block is crucial for good performance under \gls{ae} decoding \cite{Stuttgart_AE_PC,Paris_AE_v1}. 
RM codes are fully symmetric codes, i.e., $t=n$, and are thus affine invariant while polar codes are usually non-symmetric, i.e., $t=1$ \cite{Paris_AE_v1}.
Partially-symmetric codes correspond to codes having a symmetry $2\leq t\leq n-1$ \cite{PSMC}.

\section{Partially-Symmetric Reed-Muller Codes }\label{sec:distribution_symmetric}
Codes known to have good performance under \gls{ae} decoding have peculiar properties. However, it is not clear if such codes actually exist for every code dimension $K$. 
In this section, the feasibility of these codes is studied for lengths $32\leq N\leq 256$.
\subsection{Definition}
The definition of \gls{rmpsc} codes is given.
\begin{definition}[Partially-Symmetric Reed-Muller code]\label{def:RMPSC}
A $\Ccal(N,K,\Gcal)$ \gls{rmpsc} code is a RM-polar code and is $t$-symmetric with $t>1$. 
\end{definition}

A \gls{rmpsc} code is compliant with the \gls{upo}.
By being $t$-symmetric, a \gls{rmpsc} code exhibits an advantageous affine automorphism group $\Acal=\BLTA(s_1,\dots,s_l=t>1)$.
The code distance is maximised at dimension $K$ since it is a RM-polar code.
Hence, a \gls{rmpsc} code combines the most important properties required for good performance under \gls{ae} decoding \cite{BPL,Stuttgart_AE_PC,Paris_AE_v1}.
Similar codes were introduced in \cite{PSMC}.
Some of the existing polar codes constructions for AE decoding \cite{Stuttgart_AE_PC,Paris_AE_v2,Perm_auto_subcode} can be described through our formalism. 
However, this is the first time that such a construction is formally defined, and its major characteristics highlighted.
The dimensions of the partial derivatives of a \gls{rmpsc} code satisfy
\begin{align}\label{eq:not_increasing}
    K_0\geq\dots\geq K_{n-1}.
\end{align}
If \eqref{eq:not_increasing} is not verified, given $j>i$ and $K_j>K_i$, the monomial set $\Gcal$ would be composed of more monomials including the variable $\Vbar_j$ than monomials including $\Vbar_i$.
This leads to a contradiction with the decreasing property of $\Gcal$.

Since the code symmetry $t$ is defined as the number of partial derivatives having the lowest dimension \cite{PSMC}, a $t$-symmetric code of length $N$ verifies:
\begin{align}\label{eq:t_dimensions}
    K_{n-t-1}>K_{n-t}=\dots=K_{n-1}.
\end{align}
\subsection{Proposed Generator of \gls{rmpsc} Codes}\label{subsec:generator}
In the following, the \gls{rmpsc} code generator with length $N$ is based on the minimal information set $\Ical_{min}^N$ defined in \eqref{eq:Imin}. 
If the same generators, according to the decimal numeric system, are used to generate a code of length $N$ and $2N$, Theorem \ref{theo:Imin_A} shows that the code of length $2N$ exhibits useful affine automorphism group properties based on the code of length $N$.
\begin{theorem}[Affine automorphism group of generated \gls{rmpsc} codes]\label{theo:Imin_A}
By using the same decimal generators for $\Ical_{min}^N$ and $\Ical_{min}^{2N}$, if $\Ical_{min}^N$ generates a $\Ccal(N,K,\Ical)$ RM-polar code with $\Acal(\Ccal)=\BLTA\left(s_1,s_2,\dots,s_{l}\right)$,  then $\Ical_{min}^{2N}$ generates a $\Ccal'(2N,K',\Ical')$ \gls{rmpsc} code with $\Acal(\Ccal')=\BLTA\left(s_1,s_2,\dots,s_{l}+1\right)$.
\end{theorem}
\begin{proof}
Despite having the same generators in the decimal numeric system,
within the polynomial formalism, monomial generators of $\Ccal'$ are all appended by the variable $\Vbar_n$, hence $\frac{\partial \Ccal'}{\partial \Vbar_n} = \Ccal$ and $\Acal\left(\frac{\partial \Ccal'}{\partial \Vbar_n}\right)=\BLTA\left(s_1,s_2,\dots,s_{l}\right)$.
As mentioned in \cite{few_auto}, $\Acal(\Ccal)=\Acal\left(\frac{\partial \Ccal'}{\partial \Vbar_n}\right)$ is a partition of the affine automorphism group $\Acal(\Ccal')$ on variables $\{\Vbar_0,\dots,\Vbar_{n-1}\}$ ($\Vbar_n$ excluded).
Thus, we have $\Acal(\Ccal')=\BLTA\left(s_1,s_2,\dots,s_{l}+1\right)$ or $\Acal(\Ccal')=\BLTA\left(s_1,s_2,\dots,s_{l},1\right)$.
By definition, the dimensions of the partial derivatives are not increasing, namely $K_n\leq K_{n-1}$ \eqref{eq:not_increasing}. However, by appending the variable $\Vbar_n$ to every generator in $\Ical_{min}^N$, monomials including $\Vbar_n$ are as generated as monomials including $\Vbar_{n-1}$ leading to $K_n= K_{n-1}$.
Hence, the possibility that $\Acal(\Ccal')=\BLTA\left(s_1,s_2,\dots,s_{l},1\right)$ is discarded.
As a consequence, the generated code is $s_l+1$-symmetric, i.e., $\Acal(\Ccal')=\BLTA\left(s_1,s_2,\dots,s_{l}+1\right)$.
\end{proof}
Theorem~\ref{theo:Imin_A} permits to easily generate a \gls{rmpsc} code with large symmetry by using known generators generating a \gls{rmpsc} code of length $N$.
Moreover, if $\Ical_{min}^N$ generates a non-symmetric ($t=1$) RM-polar code of length $N$, by using the same generators to construct $\Ical_{min}^{2N}$, $\Ical_{min}^{2N}$ is generating a \gls{rmpsc} code of length $2N$ of symmetry $t=2$.
%
Hence, more \gls{rmpsc} codes are expected for longer code lengths; for $N=\{32,64,128,256\}$, $22, \,83, \,515,\, \text{and}\, 4275$ \gls{rmpsc} codes have been found using a restricted number of generators, i.e., here with $\left|\Ical_{min}^N\right|\leq 3$.
Limiting the number of generators to 3 is motivated by \cite[Fig.\,3]{Stuttgart_AE_PC}, since a larger $\left|\Ical_{min}^N\right|$ would likely produce non-symmetric codes.
For $N=256$ and $\left|\Ical_{min}^{256}\right|=3$, $81.5\%$, $17.2\%$, and $1.3\%$ of the codes respectively have a symmetry of 2, 3, and 4. 
If $\left|\Ical_{min}^{256}\right|=1$, $56.5\%$ of the codes are at least $t=4$ symmetric.

\begin{proposition}[Code rate of generated \gls{rmpsc} codes]\label{prop:rate}
If the same generators, according to the decimal system, are used to generate the codes $\Ccal(N,K,\Ical_{min}^N)$ and $\Ccal'(2N,K',\Ical_{min}^{2N})$, then $\frac{K'}{2N}\geq\frac{K}{N}$.
\end{proposition}
\begin{proof}
Let us denote $\Ical$ and $\Ical'$ as the information sets of $\Ccal$ and $\Ccal'$, respectively. 
Since $\frac{\partial \Ccal'}{\partial \Vbar_n}=\Ccal$, $\Ical'$ is decomposed as $\Ical$ in the upper half (between 0 and $N-1$) and $K''$ indices are implied by the partial order in the lower half (between $N$ and $2N-1$).
The partial order implies a code with more information bits in the second part leading to $K''\geq K$ extra information bits (with $K''=K$ if $K=N$). Thus, we have $\frac{K'}{2N}=\frac{K''+K}{2N}\geq\frac{K}{N}$.
\end{proof}
Proposition~\ref{prop:rate} states that using the same set of generators for a longer code produces a code with a higher code rate.
Hence, only \gls{rmpsc} codes are expected for moderate to high code rates using Theorem~\ref{theo:Imin_A}.
As an example, for $N=\{32,64,128,256\}$, the codes generated by $\Ical_{min}^{N}=\{27\}$ have rates $\sfrac{4}{32}<\sfrac{19}{64} <\sfrac{60}{128}<\sfrac{158}{256}$ and symmetry $2<3<4<5$.
\begin{figure*}[t!]
	\begin{subfigure}{0.37\columnwidth}
	  \centering
    \includegraphics[width=0.99\columnwidth]{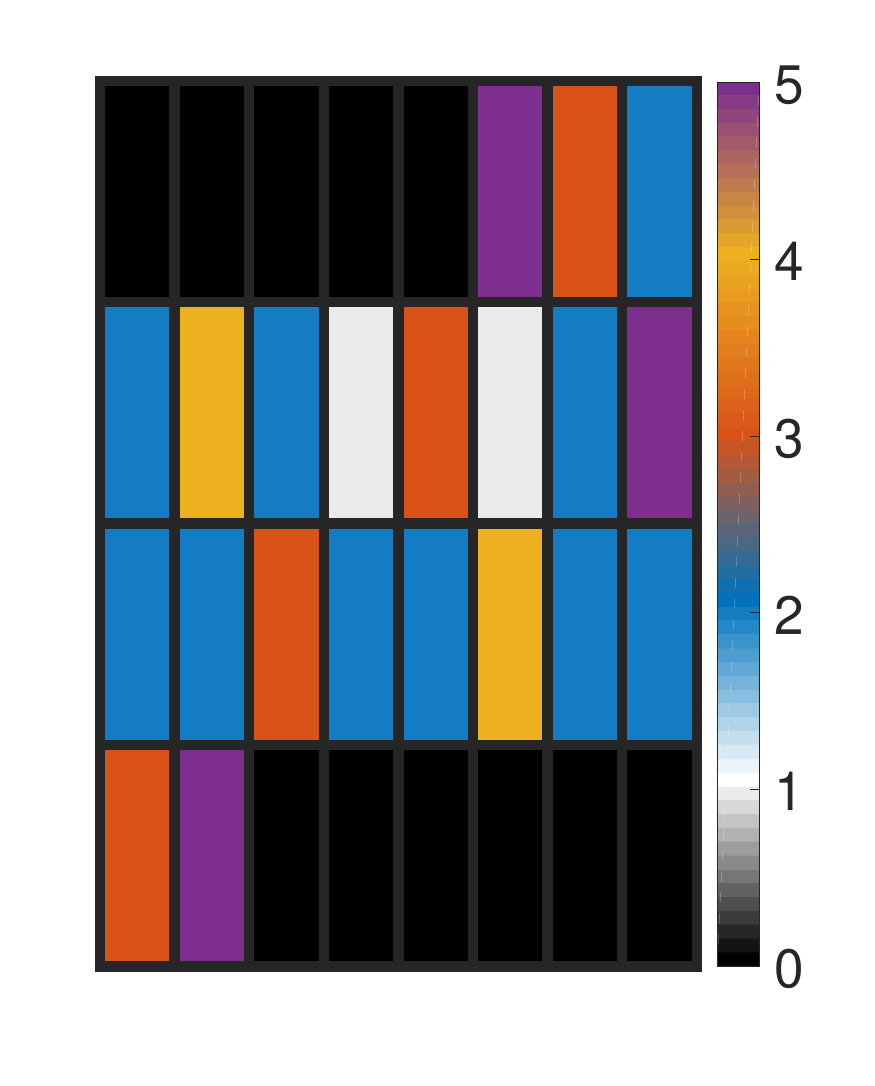}
  \caption{}
  \label{fig:s32}
  \end{subfigure}
  	\begin{subfigure}{0.37\columnwidth}
	  \centering
    \includegraphics[width=0.99\columnwidth]{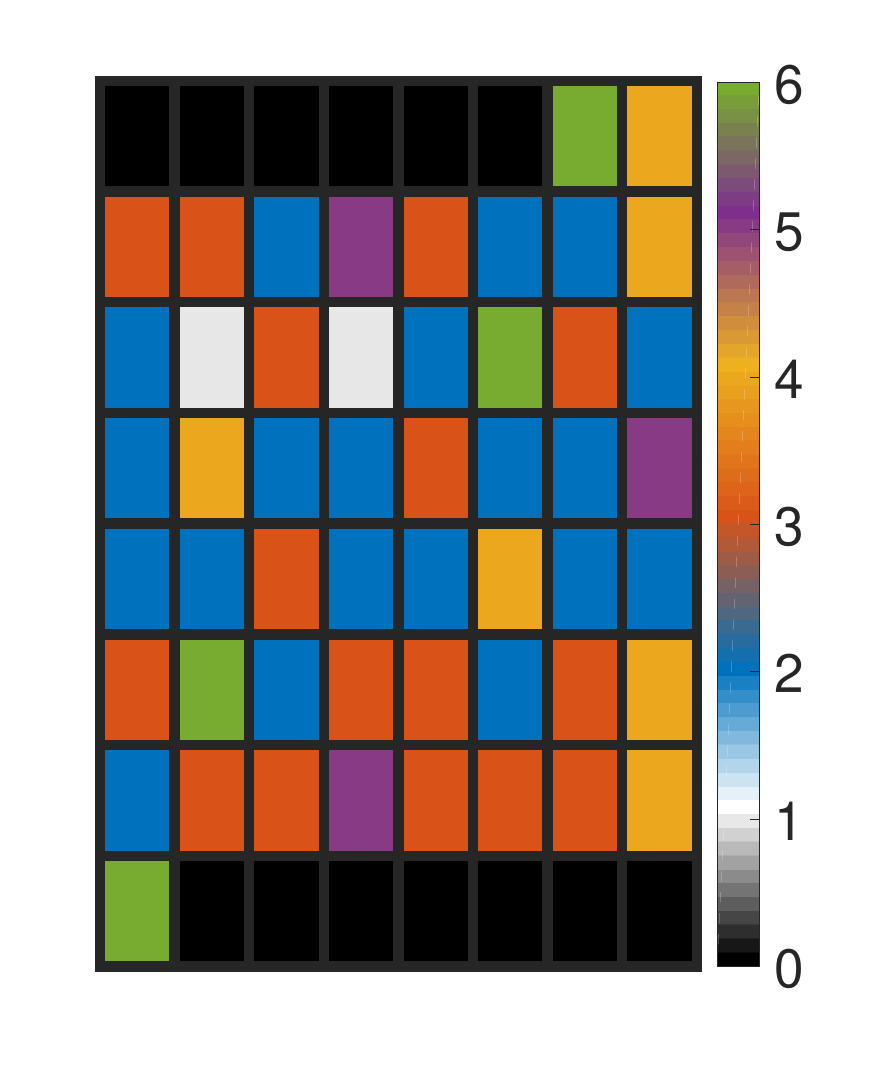}
  \caption{}
  \label{fig:s64}
  \end{subfigure}
  	\begin{subfigure}{0.63\columnwidth}
	  \centering
    \includegraphics[width=0.99\columnwidth]{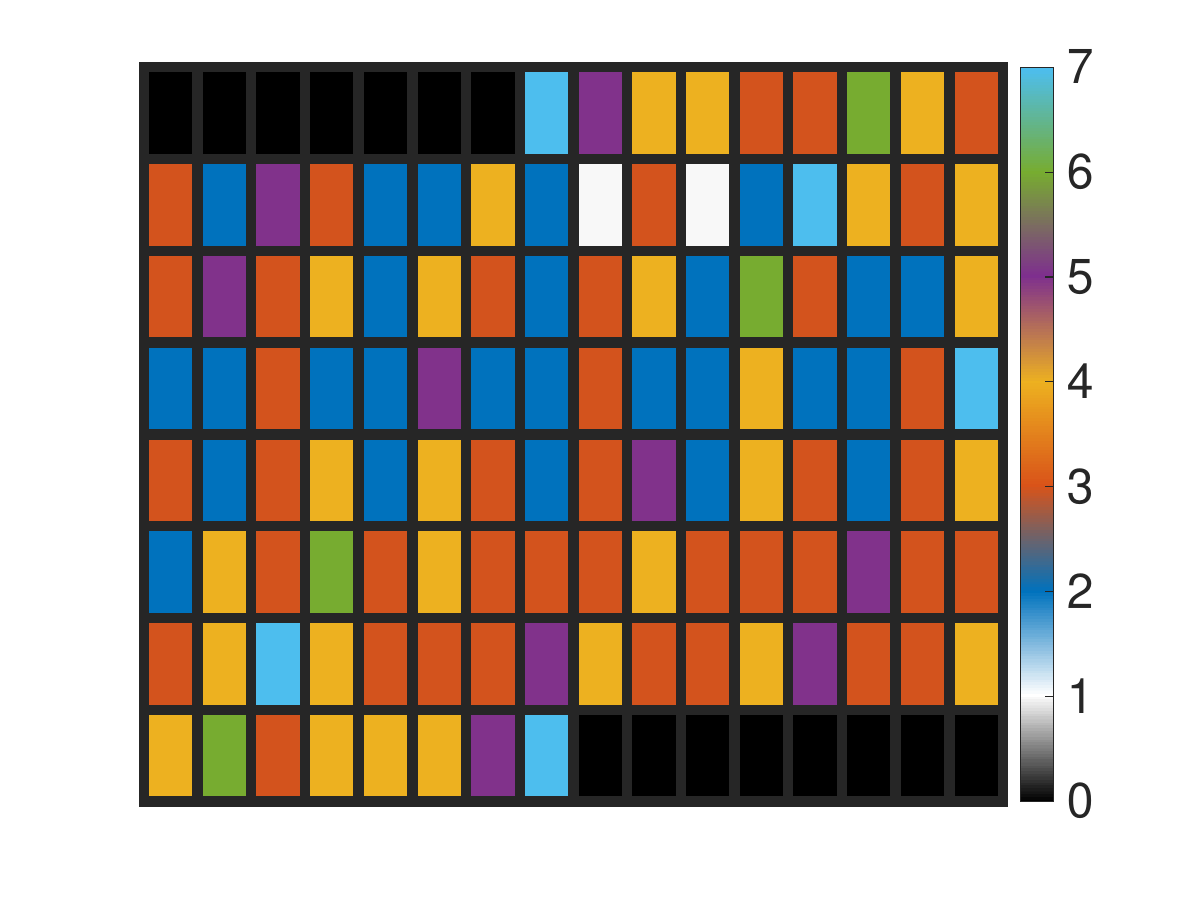}
  \caption{}
  \label{fig:s128}
  \end{subfigure}
  	\begin{subfigure}{0.63\columnwidth}
	  \centering
    \includegraphics[width=0.99\columnwidth]{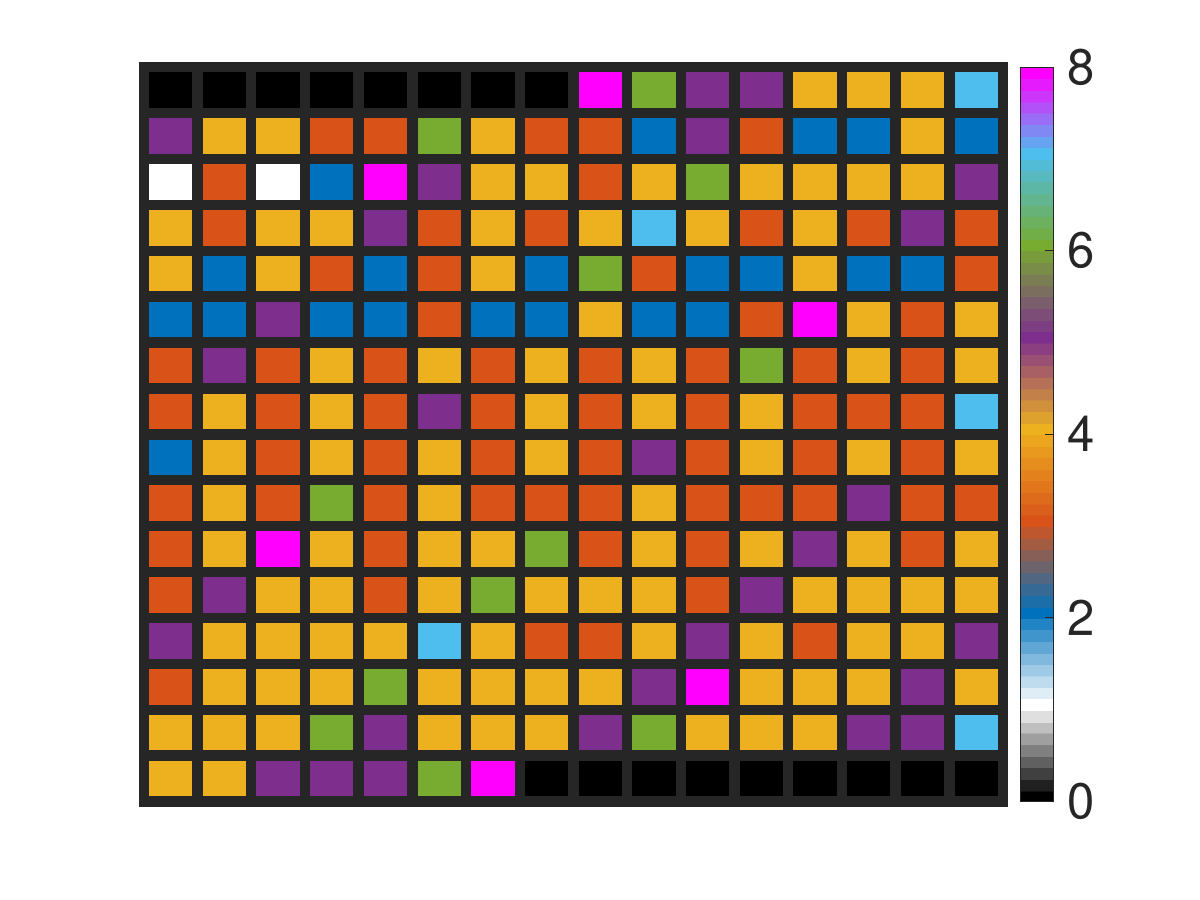}
  \caption{}
  \label{fig:s256}
  \end{subfigure}
  \caption{Maximum symmetry of \gls{rmpsc} codes for $N=$ (a) 32 (b) 64 (c) 128 (d) 256. Dimensions increase row by row starting from top left. The first and last dimensions are omitted as they have a very large absorption group $[\pid]$. }\label{fig:all}
  \end{figure*}
\subsection{Distribution of \gls{rmpsc} Codes}
Next, we investigate the \emph{distribution of \gls{rmpsc} codes}, i.e., the maximum symmetry value found for \gls{rmpsc} codes of lengths $N=\{32,64,128,256\}$.

\begin{figure}[t]
	  \centering
    \includegraphics[width=\columnwidth]{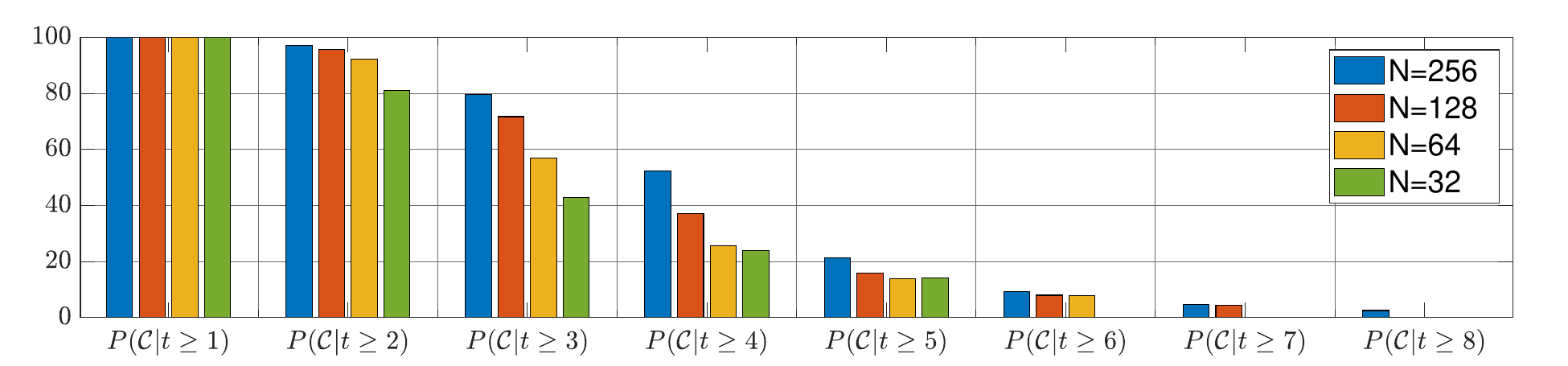}
  \caption{Percentage $P(\Ccal|t\geq x)$ of \gls{rmpsc} codes with at least symmetry $t\geq x$ for  $N=\{32,64,128,256\}$.}
  \label{fig:percentage}
 \end{figure}

\autoref{fig:all} depicts the distribution of \gls{rmpsc} codes of lengths $N=\{32,64,128,256\}$.
For all code lengths, two dimensions do not possess a single \gls{rmpsc} code, always located second-to-last and fourth-to-last before the dimension of $\mathcal{R}(2,n)$.
The existence of at least one \gls{rmpsc} code for all moderate to high code rates is verified.
\autoref{fig:percentage} shows the percentage of codes with a symmetry greater or equal to a value. It can be seen that, for a fixed symmetry, this percentage grows with $N$.
For example, for $N=256$, $52.3\%$ of the dimensions support a \gls{rmpsc} code with $t\geq 4$ whereas this percentage drops to $37.1\%,\,25.5\% \, \text{and} \,23.8\%$ for $N=\{128,\,64,\,32\}$.
As depicted in \autoref{fig:all}, a high code rate facilitates large symmetry, which is explained with \autoref{theo:Imin_A} and Proposition~\ref{prop:rate}.

The set $\Ical_{min}^N$ composed of generators having values between $0$ and $\sfrac{N}{2}-1$ is used to generate \gls{rmpsc} codes, as suggested in Section~\ref{subsec:generator}.
As expected by Proposition~\ref{prop:rate}, the low code rates are difficult to obtain with this method.
As an example, $23$ dimensions are not achievable for $N=256$, all located between $K=10$ and $K=44$.
For $N=128$, $8$ dimensions are not achievable.
For the remaining dimensions, the generator proposed in \cite{Paris_AE_v2} is used.
This method is based on code construction, but requires $\Acal(\Ccal)$ to be performed.
Since $\Acal(\Ccal)$ has many different patterns, an extensive search of codes with the desired $\Acal(\Ccal)$ is more involved than our method.

\subsection{Absorption Group $[\pid]$ of \gls{rmpsc} Codes}
\begin{table}[t]
	\centering
	\scriptsize
	\caption{Relative number of absorption groups $[\pid]$} 
	\label{tab:abs_code}
	\begin{tabular}{|c|l|c|c|c|c|}
		\cline{3-6}		
		\multicolumn{2}{c|}{}& \cellcolor{matlab5}32 & \cellcolor{matlab3}64 & \cellcolor{matlab2}128 & \cellcolor{matlab1}256\\
		\hline
		\multicolumn{1}{|c|}{\multirow{2}{*}{\footnotesize{$[\pid]=$}}} & $\Sbf_{\pid}=(1,\dots,1)$ & $1/21$& 6/52 & 22/116 & 58/251\\
		\cline{2-6}
		 &$\Sbf_{\pid}=(2,1,\dots,1)$ & 11/21& 27/52 & 62/116 & 143/251\\
		\cline{2-6}
		\multicolumn{1}{|c|}{\multirow{2}{*}{\footnotesize{$\BLTA(\Sbf_{\pid})$}}}&$\Sbf_{\pid}=(3,1,\dots,1)$ & 2/21& 4/52 & 7/116 & 9/251\\
		\cline{2-6}
		&$\mathrm{max}(\Sbf_{\pid}\setminus s'_1)>1$& 7/21& 15/52 & 25/116 & 40/251\\
		\hline
	\end{tabular}
\end{table}
\noindent Permutations in \gls{ae}-\gls{sc} must be carefully chosen. 
Some permutations are absorbed by \gls{sc} decoding resulting in no gain \cite{Stuttgart_AE_RM,SC_invariant}.
These permutations form the absorption group $[\pid]=\BLTA(\Sbf_{\pid})$ of \gls{sc} \cite{SC_invariant}, where $\Sbf_{\pid}$ denotes the profile absorbed by \gls{sc}.
$[\pid]$ depends on $\Ical$ \cite{Paris_AE_v2,SC_invariant}.
The following proposition shows a result on the \gls{sc} absorption group for most of the \gls{rmpsc} codes.

\begin{proposition}[SC absorption group of \gls{rmpsc} codes]\label{prop:abs}
A  $\Ccal(N,K,\Gcal)$ \gls{rmpsc} code with $n\leq K\leq N-n-1$, has a \gls{sc} absorption group 
\begin{align}
    [\pid]=\BLTA(\Sbf_{\pid})=\BLTA(s'_1,\dots,s'_t=1).
\end{align}
\end{proposition}
\begin{proof}
We have $s'_t>1\iff\Fcal\subseteq[N/2]$ or all information bits are in the last $\sfrac{N}{2}$ bits \cite[Alg.\,1, lines 12--23]{SC_invariant}. 
Let us prove that it is impossible for \gls{rmpsc} codes.
We divide the proof in two cases, either the monomial set $\Gcal$ includes at least one monomial including the variable $\Vbar_{n-1}$ or it does not.

In the first case, for dimensions slightly greater than $n$, i.e, close to the dimension of $\mathcal{R}(1,n)$, at least one monomial in $\Gcal$ belongs to the upper half part.
Indeed, $\mbar_{\sfrac{N}{2}-1}=\Vbar_{n-1}\in\Gcal$, thus at least one information bit is in the upper part leading to $s'_t=1$. 
For dimensions slightly smaller than $N-n-1$, i.e., close to the dimension of $\mathcal{R}(n-2,n)$, the RM-polar property ensures that $\mbar_{\sfrac{N}{2}}=\displaystyle{\prod_{i=0}^{n-2}\Vbar_i}\notin\Gcal$, hence $\Fcal\not\subseteq[N/2]$ leading to  $s'_t=1$.
For intermediate rate, the previous statement remains, lower and upper half part are mandatory composed of frozen and information bits, resulting in an absorption group with $s'_t=1$.

If no monomial in $\Gcal$ includes the variable $\Vbar_{n-1}$, \cite[Theorem 2]{Paris_AE_v1} states that the code $\Ccal$ is non-symmetric, i.e., $\Acal(\Ccal)=\BLTA(s_1,\dots,s_l=1)$. 
Hence, since $[\pid]$ is smaller than $\Acal(\Ccal)$, the code has $[\pid]=\BLTA(s'_1,\dots,s'_t=1)$.
Thus, the profile absorbed by \gls{sc} $\Sbf_{\pid}$ has $s'_t=1$ for all cases.
\end{proof}
It was shown in \cite{Paris_AE_v1} that permutations generated with ones in the lower right corner of transformation matrix $\mathbf{A}$ \eqref{eq:GA} enhance \gls{ae} decoding performance.
Thus, all \gls{rmpsc} codes exhibit valuable permutations as some of these permutations are not absorbed according to Proposition~\ref{prop:abs}.
%

If $K<n$ or  $K> N-n-1$, $\Sbf_{\pid}$ may have $s'_t> 1$ for the reasons invoked in the proof of Proposition~\ref{prop:abs}, leading to automorphisms that are not beneficial \cite{SC_invariant}.
Thus, the symmetries of these codes are not depicted in \autoref{fig:all}.


\autoref{tab:abs_code} recapitulates the statistics on the absorption groups $[\pid]$ of codes from \autoref{fig:all}.
Overall, the absorption groups are mostly composed of a single block on the top left of the $\BLTA$ structure, permuting variables of low indices together.
Such absorption groups represents $66.6\%, 71.1\%,78.5\% \, \text{and}\, 84.1\%$ of the investigated \gls{rmpsc} codes of length $32,64,128 \, \text{and}\, 256$.

\section{Simulation Results}\label{sec:results}
In this section, we present simulation results of partially-symmetric codes under \gls{ae}-$M$-\gls{sc}, denoting \gls{ae}-\gls{sc} with $M$ $\adec$ instances, transmitted with BPSK modulation over the AWGN channel.
Short and long codes are shown to illustrate the impact of \gls{rmpsc} codes under \gls{ae}-\gls{sc} decoding.
\Gls{ml} bounds are approximated using the truncated union bound.

%
\begin{figure}[t]
	  \centering
    \input{figures/128_60}
  \caption{BLER performance comparison for information rate $R=\sfrac{60}{128}$ under various construction designs.}
  \label{fig:simulation}
 \end{figure}
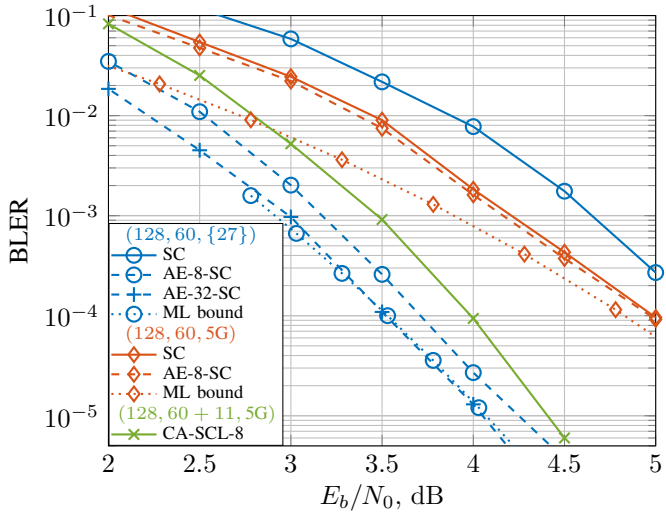
In Fig.~\ref{fig:simulation}, the $(128,\,60,\,\Ical_{min}^{128}=\{27\})$  \gls{rmpsc} code, decoded under \gls{ae}-8-\gls{sc}, outperforms by $0.25$\, dB the 5G polar code under \gls{cascl} decoding for the same list size $M=L=8$.
The \gls{rmpsc} code, having 2205 equivalence classes, reaches its \gls{ml} bound for $M=32$.
The 5G polar code, that is neither RM-polar nor partially symmetric, is not suitable for \gls{ae}-\gls{sc} having only 3 equivalence classes.

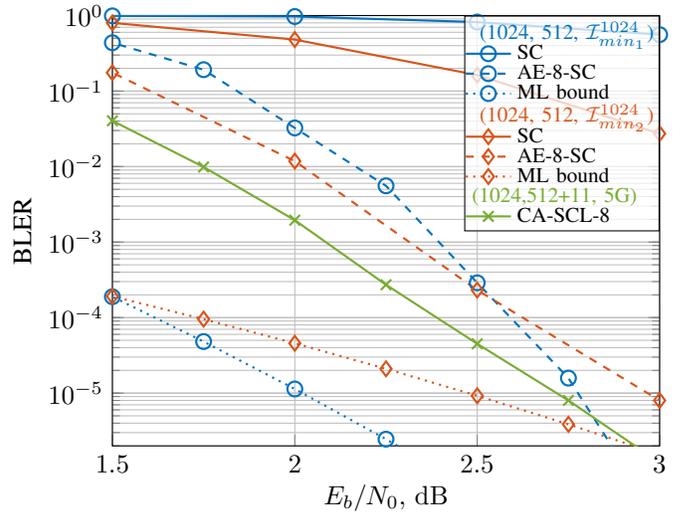
\begin{figure}[t]
  \centering
    \input{figures/1024_512}
  \caption{BLER performance comparison for information rate $R=\sfrac{512}{1024}$ under various construction designs.}
  \label{fig:simulation1024}
  \end{figure}

For $N=1024$ and $K=512$, \autoref{fig:simulation1024} shows the performance of $\Ccal_1$ defined by $\Ical^{1024}_{{min}_1}=\{63,121\}$ and $\Ccal_2$ defined by $\Ical_{{min}_2}^{1024}=\{183, 207, 241, 391, 928\}$. 
$\Ccal_1$ is retrieved with Theorem~\ref{theo:Imin_A} and is \gls{rmpsc} with symmetry $t=7$, while $\Ccal_2$, retrieved with Algorithm 1 in \cite{Paris_AE_v2}, has symmetry $t=3$ but is not a \gls{rmpsc} code.
Thus, $\Ccal_2$ has a small minimum code distance with respect to $\Ccal_1$, thus has a worse \gls{ml} bound.
However, this effect is countered by the fact that $\Ccal_2$ has only 192 minimum distance codewords, leading to an acceptable \gls{ml} bound with respect to $\Ccal_1$.
The large symmetry of $\Ccal_1$, as well as its length $N=1024$, dramatically decreases its performance under \gls{sc} \cite{PSMC}.
However, its large symmetry value allows for more diversity under \gls{ae}-\gls{sc}.
$\Ccal_1$ beats $\Ccal_2$ with an unpractical number of automorphisms $M=2048$.
Thus, \gls{rmpsc} codes may not be the best solution if \gls{ae} is applied for longer codes.
We also note the relatively poor performance of \gls{ae}-\gls{sc} for $N=1024$ with respect to \gls{cascl}, corroborating our restricted scope of code lengths studied for \gls{rmpsc} codes.

\section{Conclusions}
\label{sec:conclusions}
In this paper, the distribution of codes yielding good performance under \gls{ae} decoding was given.
These codes are RM-polar and partially symmetric.
The distribution showed that almost any dimension exhibits at least one code for  $N\leq 256$.
Moreover, the percentage of higher symmetric codes grows with code length.
We also proved that all \gls{rmpsc} codes exhibit valuable equivalence classes under \gls{sc}, confirming the existence of good codes under \gls{ae}-\gls{sc} for most dimensions.
\section*{Acknowledgement}
At the start of this work, the authors Charles Pillet and Valerio Bioglio were with Huawei Technologies France.

\balance
\bibliographystyle{IEEEbib}
\bibliography{ConfAbrv,IEEEabrv,references}

\end{document}

%% file: figures/128_60.tex
\begin{tikzpicture}

  \begin{semilogyaxis}[%
    width=\columnwidth,
    height=0.825\columnwidth,
    xmin=2, xmax=5,
    xtick={2,2.5,...,5},
    xlabel={$E_b/N_0,\,\mathrm{dB}$},
    xlabel style={yshift=0.4em},
    ymin=5e-6, ymax=1e-1,
    ylabel style={yshift=-0.1em},
    ylabel={BLER},
    yminorticks, xmajorgrids,
    ymajorgrids, yminorgrids,
    legend style={at={(0,0)},anchor=south west},
    legend style={legend columns=1, font=\scriptsize, row sep=-1mm},
    legend style={fill=white, fill opacity=1, draw opacity=1,text opacity=1}, 
    legend style={inner xsep=0.5pt, inner ysep=-1pt}, 
    legend cell align={left}, 
    mark size=1.6pt, mark options=solid,
    ]
    \addlegendimage{empty legend}
    \addlegendentry{\hspace{-5mm}\textcolor{matlab1}{$(128,60,\{27\})$}}
    
    \addplot[color=matlab1, mark=o, line width=0.8pt, mark size=2.8pt]
    table[row sep=crcr]{%
 1	0.510204\\
1.50000	0.352113\\
2	0.250627\\
2.50000	0.116550\\
3	0.0584454\\
3.50000	0.0217486\\
4	0.00776578\\
4.50000	0.00175670\\
5	0.00027\\
    };
    \addlegendentry{SC}
    
        \addplot[dashed,color=matlab1, mark=o, line width=0.8pt, mark size=2.8pt]
    table[row sep=crcr]{%
1 0.207900\\
1.50000	0.0942507\\
2 0.0348311\\
2.50000	0.0109409\\
3 0.00201861\\
3.50000	0.000258994\\
4 2.70000e-05\\
4.50000	3.700e-06\\
};
    \addlegendentry{AE-8-SC}
    
\addplot[dashed,color=matlab1, mark=+, line width=0.8pt, mark size=2.8pt]
table[row sep=crcr]{%
1 0.138122\\
1.50000	0.0532765\\
2 0.0185082\\
2.50000	0.0045045\\
3 0.000970553\\
3.50000	0.00010912\\
4 1.3e-5\\
4.50000	1e-06\\
};
    \addlegendentry{AE-32-SC}
        \addplot[dotted,color=matlab1, mark=o, line width=0.8pt, mark size=2.8pt]
    table[row sep=crcr]{%
2.78028723600244	0.00158675263374173\\
3.03028723600244	0.000664742752070708\\
3.28028723600244	0.000264892481919209\\
3.53028723600244	0.000100109085440663\\
3.78028723600244	3.57688672038788e-05\\
4.03028723600244	1.20427030880168e-05\\
4.28028723600244	3.80718199815664e-06\\
    };
    \addlegendentry{ML bound}
    \addlegendimage{empty legend}
    \addlegendentry{\hspace{-5mm}\textcolor{matlab2}{$(128,60,\text{5G})$}}
        \addplot[color=matlab2, mark=diamond, line width=0.8pt, mark size=2.8pt]
    table[row sep=crcr]{%
1	0.393701\\
1.50000	0.233645\\
2	0.121065\\
2.50000	0.0543478\\
3	0.0244738\\
3.50000	0.00900739\\
4	0.00183325\\
4.50000	0.000430834\\
5	9.6e-5\\
    };
    \addlegendentry{SC}

    \addplot[dashed,color=matlab2, mark=diamond, line width=0.8pt, mark size=2.8pt]
    table[row sep=crcr]{%
1	0.375940\\
1.50000	0.208768\\
2	0.100705\\
2.50000	0.0475059\\
3	0.0221582\\
3.50000	0.00746882\\
4	0.00161512\\
4.50000	0.000373575\\
5	9.20000e-05\\
};
    \addlegendentry{AE-8-SC}
    
    \addplot[dotted,color=matlab2, mark=diamond, line width=0.8pt, mark size=2.8pt]
    table[row sep=crcr]{%
0.280287236002435	0.261953158938647\\
0.780287236002435	0.153174967373298\\
1.28028723600244	0.0843294452346070\\
1.78028723600244	0.0433969499450641\\
2.28028723600244	0.0207063641960177\\
2.78028723600244	0.00907730940722044\\
3.28028723600244	0.00361891196296984\\
3.78028723600244	0.00129712290038050\\
4.28028723600244	0.000412639061585375\\
4.78028723600244	0.000114832336003794\\
5.28028723600244	2.75050422990110e-05\\
    };
    \addlegendentry{ML bound}

    \addlegendimage{empty legend}
    \addlegendentry{\textcolor{matlab5}{\hspace{-6mm}$(128,60+11,\text{5G})$}}
        \addplot[color=matlab5, mark=x, line width=0.8pt, mark size=2.8pt]
    table[row sep=crcr]{%
    1	0.384615\\
1.50000	0.215517\\
2	0.0826446\\
2.50000	0.0252080\\
3	0.00522766\\
3.50000	0.000915223\\
4	9.40000e-05\\
4.50000	6.00000e-06\\
    };
    \addlegendentry{CA-SCL-8}  
  \end{semilogyaxis}

\end{tikzpicture}%

    

%% file: figures/1024_512.tex
\begin{tikzpicture}

  \begin{semilogyaxis}[%
    width=\columnwidth,
    height=0.825\columnwidth,
    xmin=1.5, xmax=3,
    xtick={1,1.5,...,3},
    xlabel={$E_b/N_0,\,\mathrm{dB}$},
    xlabel style={yshift=0.4em},
    ymin=2e-6, ymax=1,
    ylabel style={yshift=-0.1em},
    ylabel={BLER},
    yminorticks, xmajorgrids,
    ymajorgrids, yminorgrids,
    legend style={at={(1,1)},anchor=north east},
    legend style={legend columns=1, font=\footnotesize, row sep=-1.2mm},
    legend style={fill=white, fill opacity=0.75, draw opacity=1,text opacity=1}, 
    legend style={inner xsep=0pt, inner ysep=0pt}, 
    legend cell align={left}, 
    mark size=1.6pt, mark options=solid,
    ]
    \addlegendimage{empty legend}
    \addlegendentry{\hspace{-5mm}\textcolor{matlab1}{(1024, 512, $\Ical_{{min}_1}^{1024}$)}}
    
    \addplot[color=matlab1, mark=o, line width=0.8pt, mark size=2.8pt]
    table[row sep=crcr]{%
1	1\\
1.50000	0.990099\\
2	0.970874\\
2.50000	0.819672\\
3	0.558659\\
    };
    \addlegendentry{SC}
    
        \addplot[dashed,color=matlab1, mark=o, line width=0.8pt, mark size=2.8pt]
    table[row sep=crcr]{%
1	0.840336\\
1.25000	0.719424\\
1.50000	0.436681\\
1.75000	0.191939\\
2	0.0324886\\
2.25000	0.00556143\\
2.50000	0.000288802\\
2.75000	1.58000e-05\\
3	2.00000e-07\\
};
    \addlegendentry{AE-8-SC}
        \addplot[dotted,color=matlab1, mark=o, line width=0.8pt, mark size=2.8pt]
    table[row sep=crcr]{%
0	0.163813422051067\\
0.250000	0.0617698596649587\\
0.500000	0.0220177961232994\\
0.750000	0.00739429919638656\\
1	0.00233139720928458\\
1.25000	0.000687560566020251\\
1.50000	0.000188914060614160\\
1.75000	4.81567963725731e-05\\
2	1.13387280452986e-05\\
2.25000	2.45439088203103e-06\\
2.50000	4.85997359256957e-07\\
2.75000	8.75678777315897e-08\\
3	1.42774489343976e-08\\
3.25000	2.09402363649984e-09\\
3.50000	2.74546333523114e-10\\
3.75000	3.19646066081422e-11\\
4	3.28162022585651e-12\\
    };
    \addlegendentry{ML bound}
    \addlegendimage{empty legend}
    \addlegendentry{\hspace{-5mm}\textcolor{matlab2}{(1024, 512, $\Ical_{{min}_2}^{1024}$)}}
        \addplot[color=matlab2, mark=diamond, line width=0.8pt, mark size=2.8pt]
    table[row sep=crcr]{%
1	0.961538\\
1.50000	0.800000\\
2	0.480769\\
2.50000	0.162866\\
3	0.0273000\\
    };
    \addlegendentry{SC}

    \addplot[dashed,color=matlab2, mark=diamond, line width=0.8pt, mark size=2.8pt]
    table[row sep=crcr]{%
0	1\\
0.500000	0.980392\\
1	0.645161\\
1.50000	0.175131\\
2	0.0118092\\
2.50000	0.000231327\\
3	8.00000e-06\\
};
    \addlegendentry{AE-8-SC}
    
    \addplot[dotted,color=matlab2, mark=diamond, line width=0.8pt, mark size=2.8pt]
    table[row sep=crcr]{%
1	0.000689952269104880\\
1.25000	0.000369941956660541\\
1.50000	0.000191445353452283\\
1.75000	9.54211471578501e-05\\
2	4.57057565235091e-05\\
2.25000	2.09896105752880e-05\\
2.50000	9.21860644149154e-06\\
2.75000	3.86197677310506e-06\\
3	1.53895414332148e-06\\
3.25000	5.81606887091723e-07\\
3.50000	2.07807507357320e-07\\
3.75000	6.99648690661832e-08\\
4	2.21187044192107e-08\\
    };
    \addlegendentry{ML bound}

    \addlegendimage{empty legend}
    \addlegendentry{\textcolor{matlab5}{\hspace{-6mm}(1024,512+11, 5G)}}
        \addplot[color=matlab5, mark=x, line width=0.8pt, mark size=2.8pt]
    table[row sep=crcr]{%
1	0.347222\\
1.25000	0.158562\\
1.50000	0.0404313\\
1.75000	0.00988663\\
2	0.00195051\\
2.25000	0.000272212\\
2.50000	4.50000e-05\\
2.75000	8.00000e-06\\
3	1.20000e-06\\
    };
    \addlegendentry{CA-SCL-8}  
  \end{semilogyaxis}

\end{tikzpicture}%

    